\documentstyle{l-aa}

\newcommand{\kms}{km\,s$^{-1}$}

\newcommand{\II}{{\sc ii}}

\begin{document}
\thesaurus {01 (11.01.1, 11.05.2, 11.06.1, 11.09.4, 13.09.4, 13.19.1)}

\title {No C$^+$ emission from the $z=3.137$ damped Lyman-$\alpha$ absorber
        towards PC\,1643$+$4631A}

\author {R.\ J.\ Ivison,\inst{1}  A.\ P.\ Harrison\inst{2} \and
I.\ M.\ Coulson\inst{3}}

\institute{
Institute for Astronomy, University Of Edinburgh, Blackford Hill,
Edinburgh EH9 3HJ, UK
\and
MRAO, Cavendish Laboratory, Madingley Road, Cambridge CB3 0HE, UK
\and
Joint Astronomy Centre, 660 North A'oh\=ok\=u Place, University
Park, Hilo, HI 96720, USA}

\date{Received date..............; accepted date................}

 \maketitle

 \markboth{R.\ J.\ Ivison, A.\ P.\ Harrison \& I.\ M.\ Coulson:
 No C$^+$ emission towards PC\,1643$+$4631A}{}

 \begin{abstract} We describe a search for redshifted [C\,\II] in a
                  $z=3.137$ damped Ly\,$\alpha$ absorption system that
                  has a large neutral hydrogen column density and
                  which was controversially reported to be a source of
                  CO emission, indicative of rapid star-formation
                  (Frayer, Brown \& Vanden Bout 1994; Braine, Downes
                  \& Guilloteau 1996). There is no sign of [C\,\II]
                  emission in our spectrum, which was obtained during
                  excellent observing conditions at the James Clerk
                  Maxwell Telescope (JCMT) and covers 1890\,\kms. The
                  upper limit we have placed on the integrated line
                  intensity ($3\sigma(T_{\rm MB}) < 5.9$\,K\,\kms\ for
                  a profile akin to that of the CO lines) constrains
                  the [C\,\II]/CO(1--0) line-intensity ratio to
                  $3\sigma < 8300$, based on the line intensity
                  reported by Frayer et al.\ (1994), or to $3\sigma <
                  58700$ based on the data obtained by Braine et al.\
                  (1996). These limits are consistent with values
                  measured in the Galactic plane and for nearby
                  starburst nuclei; the former, however, is
                  significantly lower than the ratio found in
                  low-metallicity systems such as the Large Magellanic
                  Cloud (which might be expected to have much in
                  common with a damped Ly\,$\alpha$ absorption system
                  at high redshift). This can be taken as evidence
                  against the reality of the CO line detections, with
                  the proviso that a system significantly larger than
                  present-day disk galaxies would {\em not} have been
                  fully covered by our small beam whereas it {\em
                  would} have been properly sampled by the Frayer et
                  al.\ observations. Finally, we demonstate (as did
                  Ivison et al.\ 1996) that knitting together
                  overlapping bands can generate erroneous results ---
                  specifically, an emission feature that has a width,
                  profile and central velocity consistent with the
                  controversial CO emission lines and which could have
                  drawn us to entirely the wrong
                  conclusions. \end{abstract}

\section{Introduction}

It has been suggested that damped Lyman $\alpha$ absorption systems
(DLAAS) --- massive clouds of gas which produce saturated Lyman $\alpha$
absorption features in the spectra of background quasars --- are the
high-redshift progenitors of current disk galaxies (Wolfe 1993).  The
properties of DLAAS are therefore interesting in terms of the
formation process of galaxies such as our own (Lanzetta et al.\ 1991).

There is some evidence that the number or size of DLAAS evolves over
the redshift range $3.0 < z < 3.5$ (White, Kinney \& Becker 1993),
which implies the conversion of gas into stars at $z \sim 3$; indeed,
some DLAAS at $z \sim 2$ have metallicities of 10--20 per cent Solar
(Pettini et al.\ 1994), and some contain sufficient dust to slightly
redden their background quasars (Pei et al.\ 1991). Although their
interstellar medium has clearly been enriched, direct evidence of star
formation has yet to be demonstrated (Hu et al.\ 1993).

If $z \sim 3$ DLAAS are protogalactic systems then they should have
reservoirs of low-metallicity gas and high star-formation rates. In
support of this, Frayer et al.\ (1994) reported CO(1--0) and (more
tentatively) CO(3--2) emission from a $z=3.137$ DLAAS towards the
quasar PC\,1643$+$4631A (Schneider, Schmidt \& Gunn 1991); their data
were consistent with the gas being clumped, with dimensions similar to
those of a galactic disk and a mass of around 10$^{12}$\,M$_{\odot}$
(we assume $q_0 = 0.5$, $H_0 = 50$\,\kms\,Mpc$^{-1}$ throughout this
paper).  The CO luminosity was estimated to be several orders of
magnitude greater than that of the Milky Way.

Braine et al.\ (1996) attempted to confirm the Frayer et al.\ result
using the IRAM interferometer, but concluded that the earlier CO
detections were spurious. They pointed out that an interferometer is
less prone to distorted baselines than the conventional single-dish
approach, and noted that its ideal application is the detection of
broad, weak lines from sources smaller than the primary beam (in this
case, $50''$ at 3\,mm). It is worth mentioning, however, that weak
sources spread over $\ge 20''$ ($\ge 35$\,kpc) would be heavily
resolved by the IRAM interferometer and hence very difficult to
detect. The NRAO 140-ft and 12-m dishes used by Frayer et al.\ have
HPBWs of $63''$ and $75''$ ($115-135$\,kpc), so the CO detections of
Frayer et al.\ can be understood (in the context of the Braine et al.\
data) only if the molecular gas proves to be extended on scales an
order of magnitude larger than the Milky Way. (CO has also been
reported towards PKS\,0528$-$250 in a DLAAS at $z = 2.14$ --- Brown \&
Vanden Bout 1992 --- and, again, the validity of the detection was
disputed --- Wiklind \& Combes 1994).

In a galactic environment, skins of atomic gas are thought to cover
each clump of molecular gas, with the thickness of the skin determined
by the abundance of dust, which provides protection from the global UV
field for the CO molecules within. In the atomic skin, incident UV
photons left over from dissociating and ionizing are absorbed by dust
grains, which cool via far-IR continuum emission; photoelectrons,
ejected as a consequence of UV absorption by the grains, heat the
atomic and molecular hydrogen. The C$^+$ ions are then collisionally
excited and emit the [C\,\II] ($^2P_{3/2}$ $\rightarrow$ $^2P_{1/2}$)
fine-structure line (Hollenbach, Takahashi \& Tielens 1991; Mochizuki
et al.\ 1994). The [C\,\II] emission is an important cooling process
in galaxies, accounting for up to 1 per cent of the far-IR luminosity
(Stacey et al.\ 1991).

Our objective here was to search for [C\,\II] --- an unmistakable
signature of star-formation activity --- in a DLAAS with a large
column density of neutral hydrogen, $N({\rm H}\,{\sc i}) = 5
\times 10^{20}$\,cm$^{-2}$ (White et al.\ 1993), and where CO had
apparently been detected, and to thereby independently confirm that
star formation is ongoing in that system and that enriched gas is
present. Demonstrating the potential of [C\,\II] as a probe of
metallicity, of DLAAS star-formation history, and of the evolution of
galaxies at these redshifts, would represent a major advance in our
studies of the early Universe.

For nearby galaxies, the rest frequency of the [C\,\II] transition
(1.900537\,THz) means that the line is inaccessible from the ground
and, to date, all detections in the near Universe have been made by
the balloon-borne experiments (Mochizuchi et al.\ 1994), the Kuiper
Airbourne Observatory (Stacey et al.\ 1991) or the Cosmic Background
Explorer (Bennett et al.\ 1994). However, at high redshifts the
[C\,\II] line is shifted into windows observable from Mauna Kea with
the 15-m JCMT; specifically, for $4.14 < z < 5.33$ the line appears in
the B window, and for $2.76 < z < 3.22$ it appears in the C window. To
date, there have been no detections of highly redshifted [C\,\II]
(e.g.\ Isaak et al.\ 1994).

\section{Observations and data reduction}

The data reported here were obtained during excellent observing
conditions in 1994 Dec and 1995 Jan, mostly as part of an experiment
in flexibly scheduled service observing at the JCMT. We used the
single-channel SIS receiver, C2, with a broad-band digital
autocorrelation spectrometer (DAS) as the backend. The beam (11$''$
FWHM, or around 20\,kpc at $z=3.137$) was nutated by 60$''$ in
azimuth, at a rate of 1\,Hz, with the telescope position-switching by
the same distance every 30\,s to alternate the signal and reference
beams. The maximum pointing offset during the observations was less
than 4$''$. $T_{\rm sys}$ ranged from 1050 to 2300\,K, and the sky
transparency was excellent. A total of 105\,min was spent on source,
with a 175 per cent overhead for sky subtraction, position switching,
calibration and pointing checks.

 \begin{table}
 \caption{Log of observations towards PC\,1643+4631A.}
 \begin{tabular}{lccc}
Mean UT date   &$T_{\rm sys}$&Central $v_{\rm lsr}$ for&Integration\\
               &/K           &$z=3.137$ /km\,s$^{-1}$  &time /s    \\
               &             &                         &           \\
1994 Dec 23.78 &      1148   &$+$0                     &3600       \\
1995 Jan 20.73 &      2305   &$-$430                   &1800       \\
1995 Jan 20.78 &      1706   &$+$430                   &1800       \\
1995 Jan 21.70 &      1396   &$-$860                   &3600       \\
1995 Jan 21.78 &      1716   &$-$430                   &1800       \\
 \end{tabular}
 \end{table}

The maximum bandwidth of the DAS (920\,MHz, or 600\,\kms\ at
459.399807\,GHz) is barely sufficient when searching for
high-frequency lines, particularly when the target line is broad as
was expected to be the case here. For this reason, four slightly
overlapping spectra were obtained (with band centres at 458.74362,
459.39987, 460.05612 and 460.71237\,GHz --- see Table~1), giving a
full velocity coverage of 1890\,\kms. These were reduced, using SPECX
V6.7 (Padman 1993), baseline-subtracted (zero order baselines) and
binned to give a velocity resolution of 40\,\kms.  The overlap regions
were averaged.

Our conversion from atmosphere-corrected antenna temperatures,
$T_A^*$, to the $T_R^*$ scale assumes a forward spillover efficiency,
$\eta_{\rm fss}$, of 70 per cent, and is accurate to $\pm10$ per
cent. The conversion factor between the $T_R^*$ scale and flux density
is $S_{\nu} = 26 \times (T_R^*/{\sc k})$\,Jy, which assumes an
aperture efficiency of 42 per cent. The beam efficiency at 460\,GHz,
measured on Mars, was $53 \pm 5$ per cent, so $T_{\rm MB} = 1.89
\times T_R^*$, with an overall uncertainty of around 20 per cent.

\section{Results}

In Fig.~1 we present our JCMT spectrum, together with the CO data from
Frayer et al.\ (1994). The overall rms is $\sigma(T_{\rm MB}) =
15$\,m{\sc k}, or $12$\,m{\sc k} if we limit ourselves to the central
portion of the spectrum. The resulting upper limit on the integrated
line intensity, calculated using the formulae derived by Seaquist,
Ivison \& Hall (1995), assuming a rectangular profile with FWHM
680\,\kms\ (similar that of the CO(1--0) line reported by Frayer et
al.\ 1994), is $3\sigma(T_{\rm MB}) < 5.9$\,K\,\kms. Frayer et al.\
(1994) give $T_{\rm MB}$(CO(1--0)) = $3.2 \pm 1.2$\,K\,\kms, or
$10.0$\,Jy\,\kms, or $9.3 \times 10^{-21}$\,W\,m$^{-2}$, hence the
measured [C\,\II]/CO(1--0) intensity ratio is $3\sigma < 8300$.

Braine et al.\ (1996) reported an rms of 1.25\,mJy in channels of
width 224\,\kms\ for their observations of CO(3--2). This translates
into a limit on the integrated line intensity of $3\sigma <
1.6$\,Jy\,\kms\ or $3\sigma < 4.4 \times 10^{-21}$\,W\,m$^{-2}$
(assuming a rectangular profile with FWHM 800\,\kms, as reported by
Frayer et al.\ 1994) --- a factor of 7.1 lower than the integrated
CO(3--2) line intensity ($11.4 \pm 3.5$\,Jy\,\kms) reported by Frayer
et al. (1994). Assuming that the CO(1--0) has been similarly
overestimated, this yields a [C\,\II]/CO(1--0) intensity ratio of
$3\sigma < 58700$.

    \begin{figure} \setlength{\unitlength}{1mm}
    \begin{picture}(80,142) \put(0,0){\includegraphics{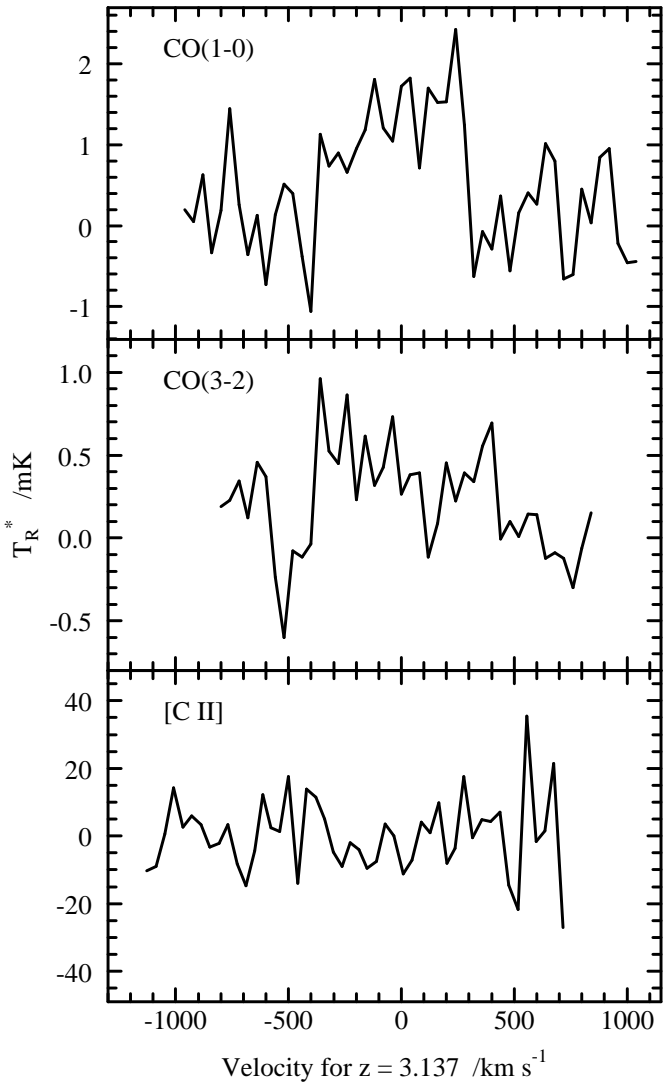}}
    \end{picture} \caption{Spectra of the $z=3.137$ DLAAS towards
    PC1643$+$4631A (this paper; Frayer et al.\ 1994). The zero point
    of the velocity scale represents the expected position of the
    plotted lines for $z=3.137$. Top: NRAO 140-ft CO(1--0); middle:
    NRAO 12-m CO(3--2); bottom: JCMT 15-m [C\,\II], with $v_{\rm
    lsr}=0$\,\kms\ corresponding to 459.399807\,GHz. The spectra have
    been binned to 40\,\kms\ in all cases. Zero-order baseline
    corrections have been applied.}
    \end{figure}

    \begin{figure}
    \setlength{\unitlength}{1mm}
    \begin{picture}(80,105)
    \put(0,0){\includegraphics{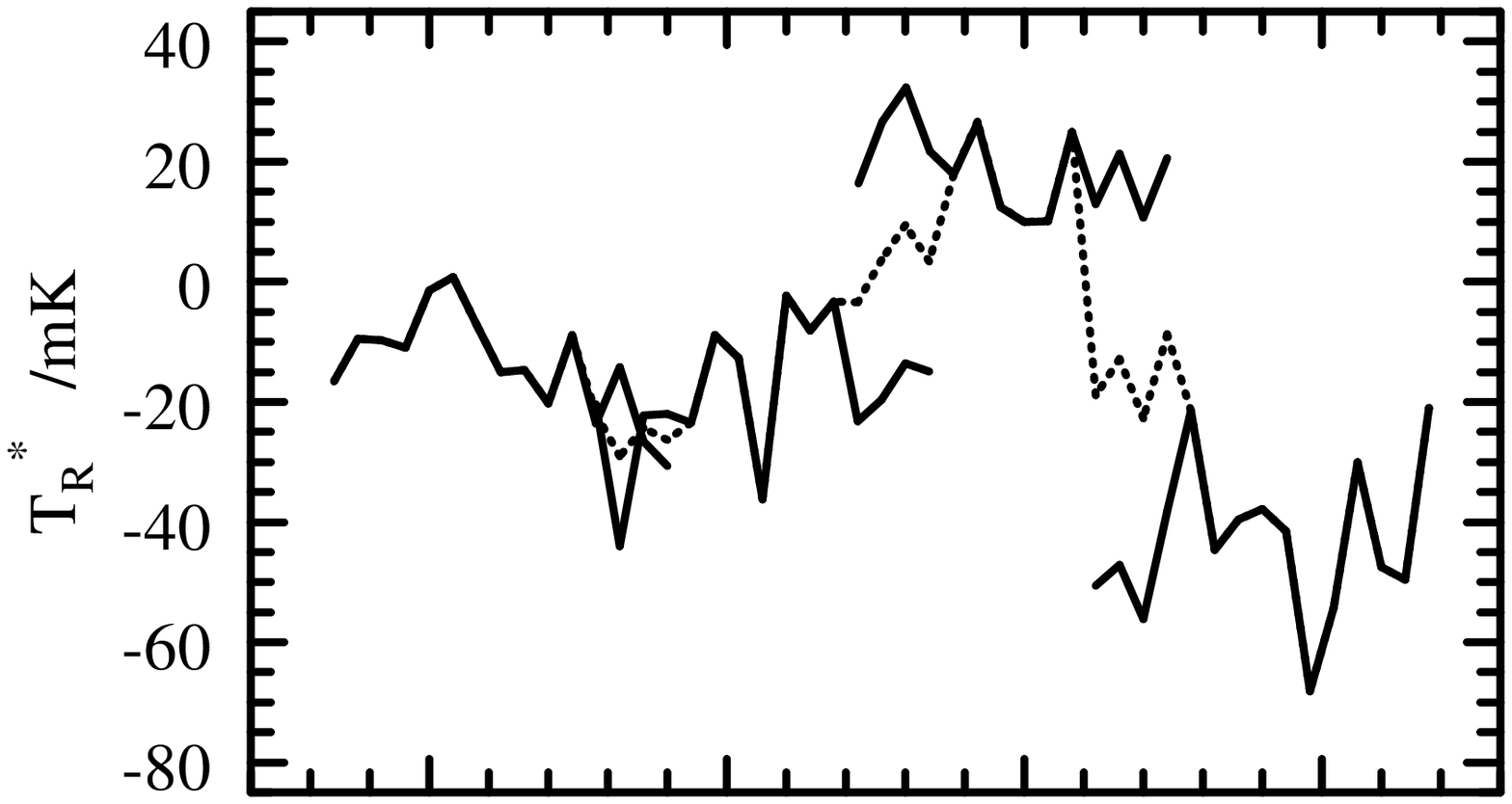}}
    \put(0,0){\includegraphics{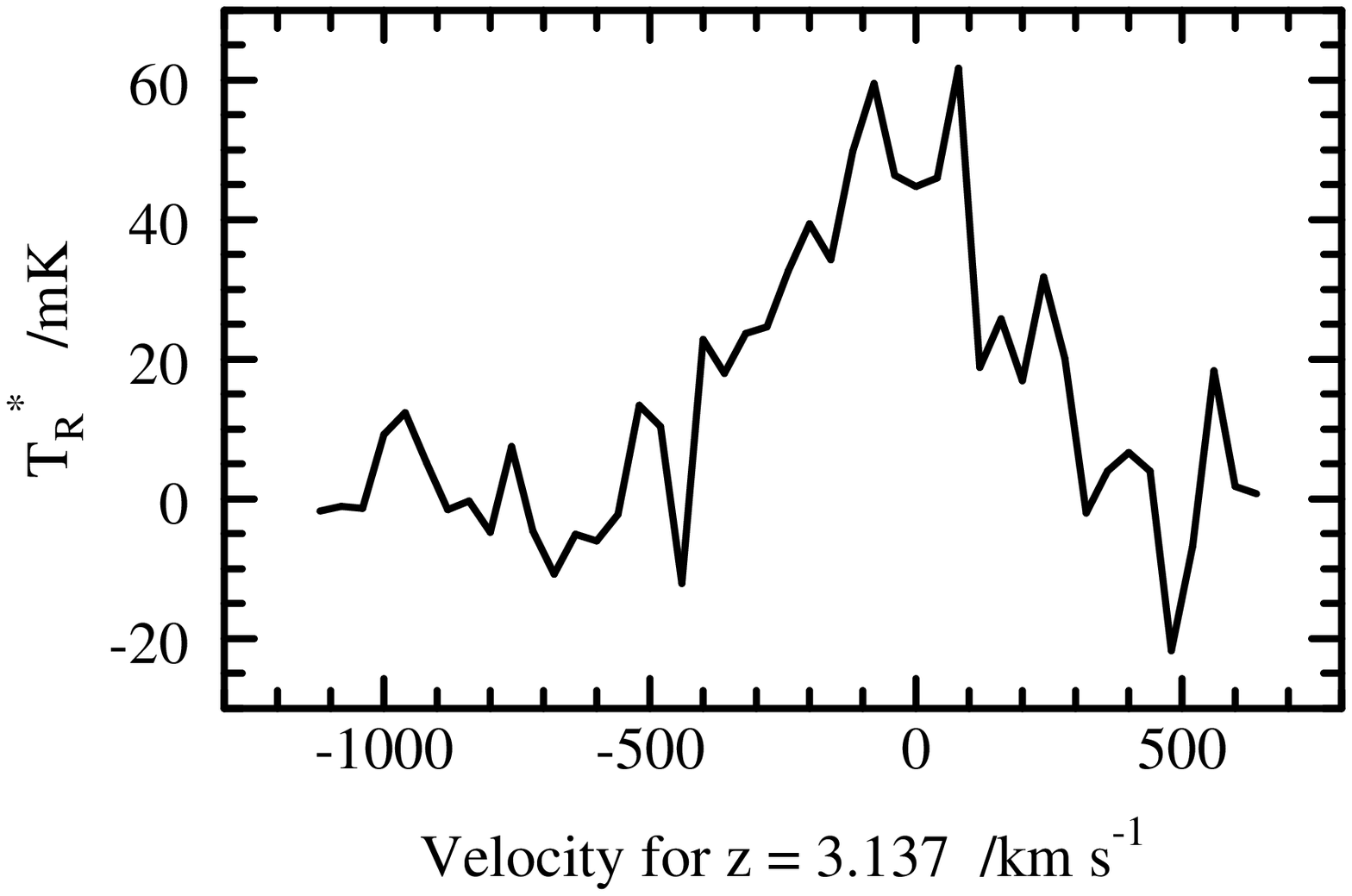}}
    \end{picture}
    \caption{Top: Spectrum ({\em dashed line}) of the DLAAS
towards PC1643$+$4631A, with $v_{\rm lsr}=0$\,\kms\ corresponding to
459.399807\,GHz ([C\,\II] for $z=3.137$), constructed from individual
600-\kms\ segments ({\em solid lines}) without correcting their baselines
(following the method of Frayer et al.\ 1994). Bottom: The combined
spectrum after subtracting a linear baseline.}
    \end{figure}

Fig.~2 shows a graphic demonstration of the dangers of coadding
overlapping spectra to improve velocity coverage. In this case,
baselines were not subtracted from the individual segments. The effect
of combining the poor baselines is to generate a very convincing
emission feature (still more so if we subtract a linear baseline at
this stage --- see the lower panel of Fig.~2). The apparent emission
line is centred at $v_{\rm lsr} = 0$\,\kms\ for $z=3.137$ and has a
full width similar to that of the controversial CO lines, though more
Gaussian in profile.

The integrated intensity of the apparent line is $45 \pm 10$\,K\,\kms\
on the $T_{\rm MB}$ scale, which would have indicated a
[C\,\II]/CO(1--0) intensity ratio of 63000. This would have led us to
completely the wrong conclusion since this value is consistent with
those of low-metallicity systems (see the discussion that
follows). Moreover, this would have been regarded as strong support
for the validity of the CO detections and as indicative of rapid
ongoing star formation in the DLAAS towards PC\,1643+4631A.

Offsets such as those seen in the upper panel of Fig.~2 are usually
the result of incomplete sky subtraction, or poor instrumental
stability. We suspect the former in this case, even though the spectra
were obtained during excellent and seemingly quite stable
conditions. It is possible that more frequent nodding between the
signal and reference beams would have reduced the offsets, but such
anomalies are a fact of life in the submillimetre regime and we can be
grateful to some extent that the baselines produced RxC2 and DAS are
such good approximations of zeroth order. There are no fool-proof
methods of achieving perfect sky subtraction and if there is a lesson
to be learned, it is that high-bandwidth receivers and spectrometers
are extremely desirable in this field.

\section{Discussion}

In nearby starbursts, the photodissociated gas represents a
substantial fraction (40 per cent) of the total gas mass in the
nuclei, and the line-intensity ratio of [C\,\II] to CO(1--0) is 4100,
with a very small scatter (Stacey et al.\ 1991). In the Galactic
plane, the ratio is around 1300 (Nakagawa et al.\ 1993), whilst in
low-metallicity regions such as 30~Dor, or the Large Magellanic Cloud
(LMC) in general, the ratio is high (77000 and 23000 for 30~Dor and
the LMC, respectively) because there are few dust grains to shield the
molecular gas. The UV therefore penetrates deep inside each clump,
dissociating CO and creating a thick skin of C$^+$ ions (Mochizuki et
al.\ 1994). Note that although the metallicity is thought to have the
dominant influence on the [C\,\II]/CO(1--0) line intensity ratio, the
global UV field strength is also expected to have some effect.

Given that DLAAS at $z > 3$ are expected to be low-metallicity
systems, perhaps similar in many respects to the LMC, the data we have
presented (in particular the low limit on the [C\,\II]/CO(1--0)
intensity ratio, based on the claimed CO(1--0) detection) support the
view that the CO detections of Frayer et al.\ (1994) were
spurious. There is, however, one proviso concerning the beams used to
sample the emission region: the area of our beam (11$''$ FWHM) was
$\sim 40$ times smaller than those used by Frayer et al., and if the
emission region proves to be an order of magnitude larger than the
Milky Way then we would not only have missed the majority of the
emitting gas, but we may well have been chopping onto some of it.

\section{Concluding Remarks}

We have searched for redshifted [C\,\II] towards a $z=3.137$ damped
Ly\,$\alpha$ absorption system that has a large neutral hydrogen
column density and which was controversially reported to be a source
of CO emission, indicative of rapid star-formation. We find no sign of
[C\,\II] emission and have placed an upper limit of $3\sigma(T_{\rm
MB}) < 5.9$\,K\,\kms\ on the integrated line intensity.

This places a useful constraint on the [C\,\II]/CO(1--0)
line-intensity ratio ($3\sigma < 8300$, based on the line intensity
reported by Frayer et al.\ 1994) which is consistent with ratios
measured in normal-metallicity systems in the present-day Universe,
but is significantly lower than the ratio found in systems with low
metallicities such as we might expect to find in high-redshift damped
Lyman $\alpha$ absorption systems. We interpret this as evidence
against the reality of the CO line detections towards this system, as
long as the system is not significantly larger than present-day disk
galaxies such as the Milky Way (which would compromise our measured ratios
on the basis of disparate beam sizes).

We have also demonstated the dangers inherent in knitting together
overlapping bands to increase velocity coverage. Clearly, wide-band
receivers and backends are urgently required if we are to generate a
trustworthy database of CO, [C\,\II], etc., spectra of high-redshift
systems.

\subsection*{ACKNOWLEDGMENTS}

The JCMT is operated by the Observatories on behalf of the UK Particle
Physics and Astronomy Research Council (PPARC), the Netherlands
Organization for Scientific Research and the Canadian National
Research Council. RJI is supported by a PPARC Advanced Fellowship.

\end{document}